\documentclass[final,5p,times,twocolumn]{elsarticle}

 \usepackage{graphics}

\usepackage{amssymb}
\usepackage{hyperref}
\hypersetup{colorlinks=true,%
           linkcolor=black,%
           anchorcolor=black,%
           citecolor=black,%
           filecolor=black,%
           menucolor=black,%
           urlcolor=black}

\usepackage{subfigure}

\newcommand{\neqcm}{\ensuremath{\mathrm{n}_{\mathrm{eq}}/\mathrm{cm}^2}}
\newcommand{\mum}{\ensuremath{\mu}m}
\newcommand{\ke}{\ensuremath{\mathrm{ke}^{-}}}
\newcommand{\e}{\ensuremath{\mathrm{e}^{-}}}

\journal{Nuclear Instruments and Methods A }

\begin{document}

\begin{frontmatter}




\title{Characterization and Performance of Silicon n-in-p Pixel Detectors  for the ATLAS Upgrades}
\author{P.~Weigell\corref{cor1}\fnref{label1}}
\ead{Philipp.Weigell@mpp.mpg.de}
\cortext[cor1]{Corresponding author}
\author{M.~Beimforde \fnref{label1}}
\author{Ch.~Gallrapp \fnref{label3}}
\author{A.~La Rosa \fnref{label3}}
\author{A.~Macchiolo \fnref{label1}}
\author{R.~Nisius \fnref{label1}}
\author{H.~Pernegger \fnref{label3}}
\author{R.H.~Richter \fnref{label1,label2}}
\address[label1]{Max-Planck-Institut f\"ur Physik, F\"ohringer Ring 6, D-80805 M\"unchen, Germany}
\address[label2]{Max-Planck-Institut Halbleiterlabor, Otto Hahn Ring 6, D-81739 M\"unchen, Germany}
\address[label3]{CERN-PH, Switzerland}

\begin{abstract}
The existing ATLAS Tracker will be at its functional limit for particle fluences of 10$^{15}$\,\neqcm{} (LHC). Thus for the upgrades at smaller radii like in the case of the planned Insertable B-Layer (IBL) and for increased LHC luminosities (super LHC) the development of new structures and materials which can cope with the resulting particle fluences is needed. N-in-p silicon devices are a promising candidate for tracking detectors to achieve these goals, since they are radiation hard, cost efficient and are not type inverted after irradiation.
A n-in-p pixel production based on a MPP/HLL design and performed by CiS (Erfurt, Germany) on 300\,\mum{} thick Float-Zone material is characterised and the electrical properties of sensors and single chip modules (SCM) are presented, including noise, charge collection efficiencies, and measurements with MIPs as well as an $^{241}$Am source. The SCMs are built with sensors connected to the current the ATLAS read-out chip FE-I3. The characterisation has been performed with the ATLAS pixel read-out systems, before and after irradiation with 24\,GeV/c protons. In addition preliminary testbeam results for the tracking efficiency and charge collection, obtained with a SCM, are discussed. 
\end{abstract}


\end{frontmatter}



\section{Introduction}
\label{sec:introduction}
The present LHC accelerator will be upgraded to reach higher luminosities in two phases \cite{SLHC}. During "Phase I", starting in 2016, the aim is to reach a luminosity of around (2-3)$\cdot$10$^{34}$\,cm$^{-2}$s$^{-1}$ by some comparatively small upgrades to the machine itself and improvements to the pre-accelerators. Examples are collimation upgrades and the Linac4 installation. Later, presumably after 2020, a major upgrade effort ("Phase II") will increase the luminosity further to 5$\cdot$10$^{34}$\,cm$^{-2}$s$^{-1}$ or higher. In this scenario the innermost layers of the ATLAS vertex detector system will have to sustain very high integrated fluences of more than 10$^{16}$\,\neqcm{} (1\,MeV equiv.) \cite{Dawson}. 

The start of Phase I leads to a first upgrade of the ATLAS pixel detector, the Insertable B-Layer (IBL) \cite{IBL-TDR}, achieved through the insertion of an additional pixel layer together with a new beam pipe of smaller radius. Three main sensor technologies are under investigation as possible candidates for the IBL detectors: CVD\footnote{Chemical vapour deposition} Diamond \cite{Wedenig1999497}, silicon 3D \cite{Parker3D}, and planar sensors. In the case of planar sensors, the standard n-in-n sensors, presently used in the ATLAS
pixel detector, and the new n-in-p devices, discussed in the following, are participating in a qualification program. Additionally, a new read-out chip FE-I4 \cite{GarciaSciveres2010}, which features smaller pitch sizes of 250\,\mum\,$\times$\,50\,\mum{} and an approximately six times bigger area than the present ATLAS read-out chip FE-I3 \cite{Peric2006178} is developed for the IBL. Since the new read-out chip is not yet available for R\&D projects the FE-I3 chip is used at the moment.

\section{N-in-p Pixel Technology \& Sensor Design}
\label{sec:setup}
The feasibility to use n-in-p sensors for these upgrades is under investigation. In contrast to the current n-in-n ATLAS pixel detector design the n-in-p technology only needs processing of the wafer on the front side, which allows to reduce the production cost.

\subsection{Sensor Design}
The sensors were produced in the framework of the CERN RD50 collaboration and the ATLAS Planar Pixel Sensor Group using a design by MPP/HLL on the CiS production lines using 18 Float-Zone 4"-wafers of 300\,\mum{} thickness. The design goals of the IBL aim for inactive edges of less than 450\,\mum{} per side. 
Thus, pixel sensor designs with reduced guard-ring structures, as opposed to the 19 guard-rings of the present sensor design, have been inserted in the wafer layout. One of these additional designs has eight and one 15 guard-rings.

\begin{figure}[ht]
\centering
\includegraphics[scale=0.28,angle=180]{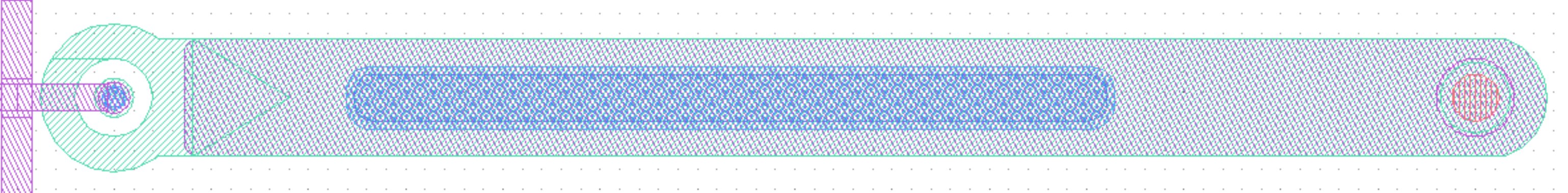}
\caption{Design of a single pixel. The metal layer is shown in pink, the implantation in green, the opening in the nitride and oxide in blue, and the opening in the passivation layer in red.}
\label{fig:Pixeldesign}
\end{figure}

Each wafer hosts ten FE-I3 compatible pixel sensors \cite{Peric2006178}, having four different kinds of pixels arranged in 18 columns and 160 rows. The ``normal'' pixels cover the central region between the second and the 17th column and extending from row zero to 152. They have an extension of 400\,\mum$\times$50\,\mum. The first and the last column consist of ``long'' pixels measuring 600\,\mum$\times$50\,\mum. The uppermost eight rows are populated alternating with ``ganged'' and ``inter-ganged'' pixels, since the last four rows are not bump bonded to the read-out chip. Therefore each ganged pixel is connected to an inter-ganged pixel and read-out via this.

Each pixel can be biased via a punch-through structure, of 20\,\mum{} diameter (cf. Fig.\,\ref{fig:Pixeldesign} on the right). Since the central dot of the n+ implantation is not directly connected to the pixel structure a reduced charge collection efficiency in this region should be expected.

In the n-in-p devices the high voltage, which is applied to the back side of the sensor, is---via the cutting edge---also present on the edges of the front side of the sensor facing the read-out chip, which is at ground potential with a distance of approximately 20\,\mum{}. A additional special passivation layer of 3\,\mum{} Benzocyclobutene (BCB) has been applied by the Fraunhofer Institut f\"ur Zuverl\"assigkeit und Mikrointegration (IZM) to prevent sparks between the sensor and the read-out chip.

\section{Readout \& System Setup}
For the read-out the current read-out chip of the ATLAS experiment (FE-I3) is used. The flip-chipping of 21 single chip modules (SCM) has been done by the IZM in three batches, of which two batches are completed.

The data was gathered using the PC-VME based TurboDAQ as well as the USBPix system. The software of the USBPix system is based partly on the ATLAS software libraries, so improvements could be used by the DAQ of ATLAS as well. Especially the threshold tuning and charge calibration algorithms are different with respect to the TurboDAQ setup and tend to yield better tunings in terms of the targeted value and its dispersion.

Source scans were conducted using a $^{90}$Sr and an $^{241}$Am source. While for the source scans with $^{90}$Sr a scintillator coupled to a photo-multiplier is used as external trigger, for the $^{241}$Am scans the internal trigger of the read-out chip is used. 

\section{Results}
\label{sec:results}
\subsection{Laboratory Characterisation}
\begin{figure}[htb]
\centering
\includegraphics[angle=90,scale=0.4]{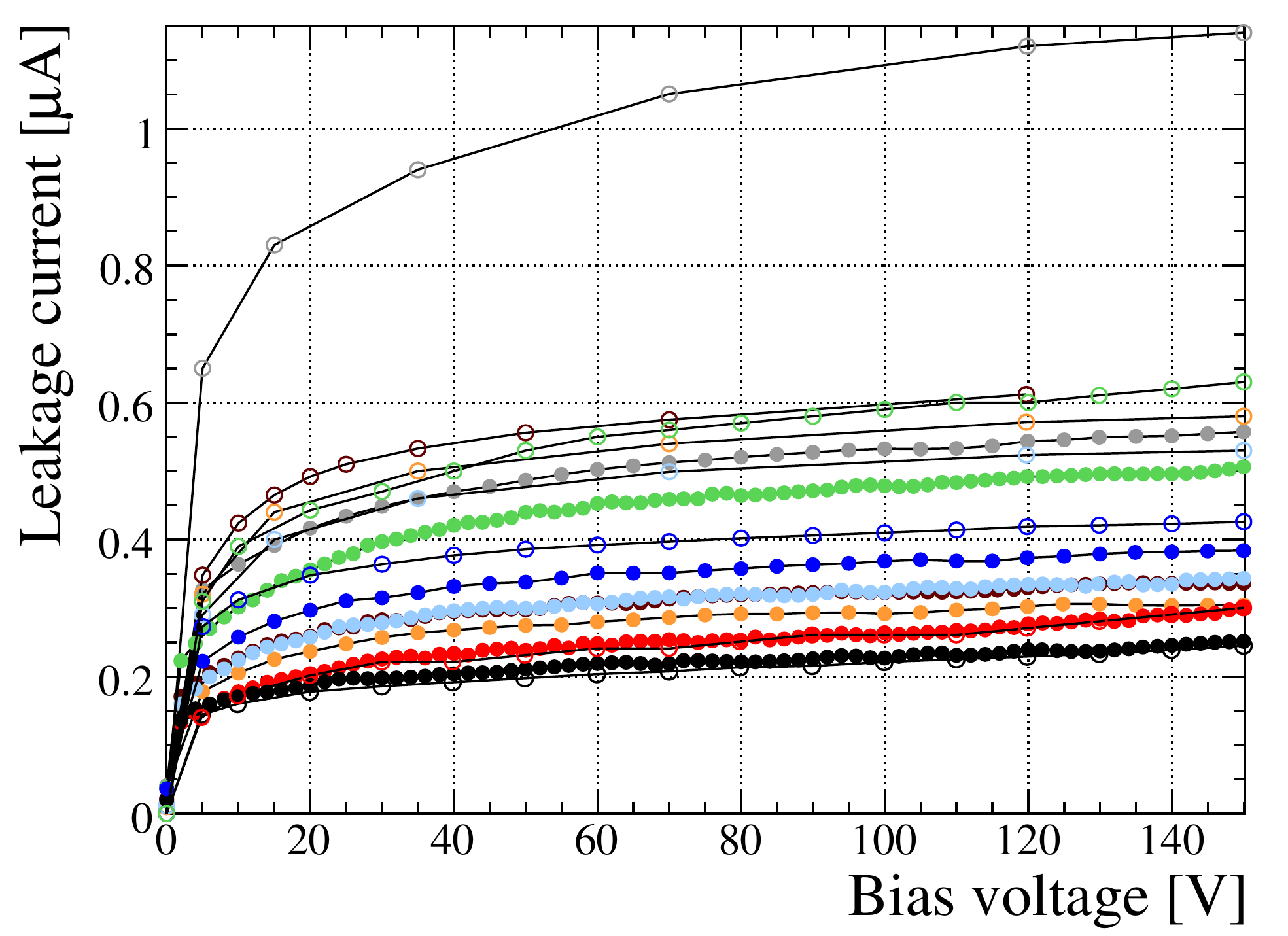}
\caption{IV characteristics of the analysed modules before (filled symbols) and after flip-chipping (open symbols) measured at room temperature. The colours indicate the different devices.}
\label{fig:FlipchipIV}
\end{figure}
All the sensors were electrically characterised before selecting the devices to be interconnected to the chip. Breakdown voltages typically in excess of 350\,V have been measured, while the depletion voltage lies at 60\,V. The IV characteristics for the selected sensors before and after flip-chipping are shown in Fig.\,\ref{fig:FlipchipIV}. With one exception leakage currents are before as well as after flip-chipping below 0.65\,$\mu\mathrm{A}$ and thus of no concern.

\subsubsection{Tuning of Threshold}
\begin{figure}[ht]
\centering
\includegraphics[angle=90,scale=0.4]{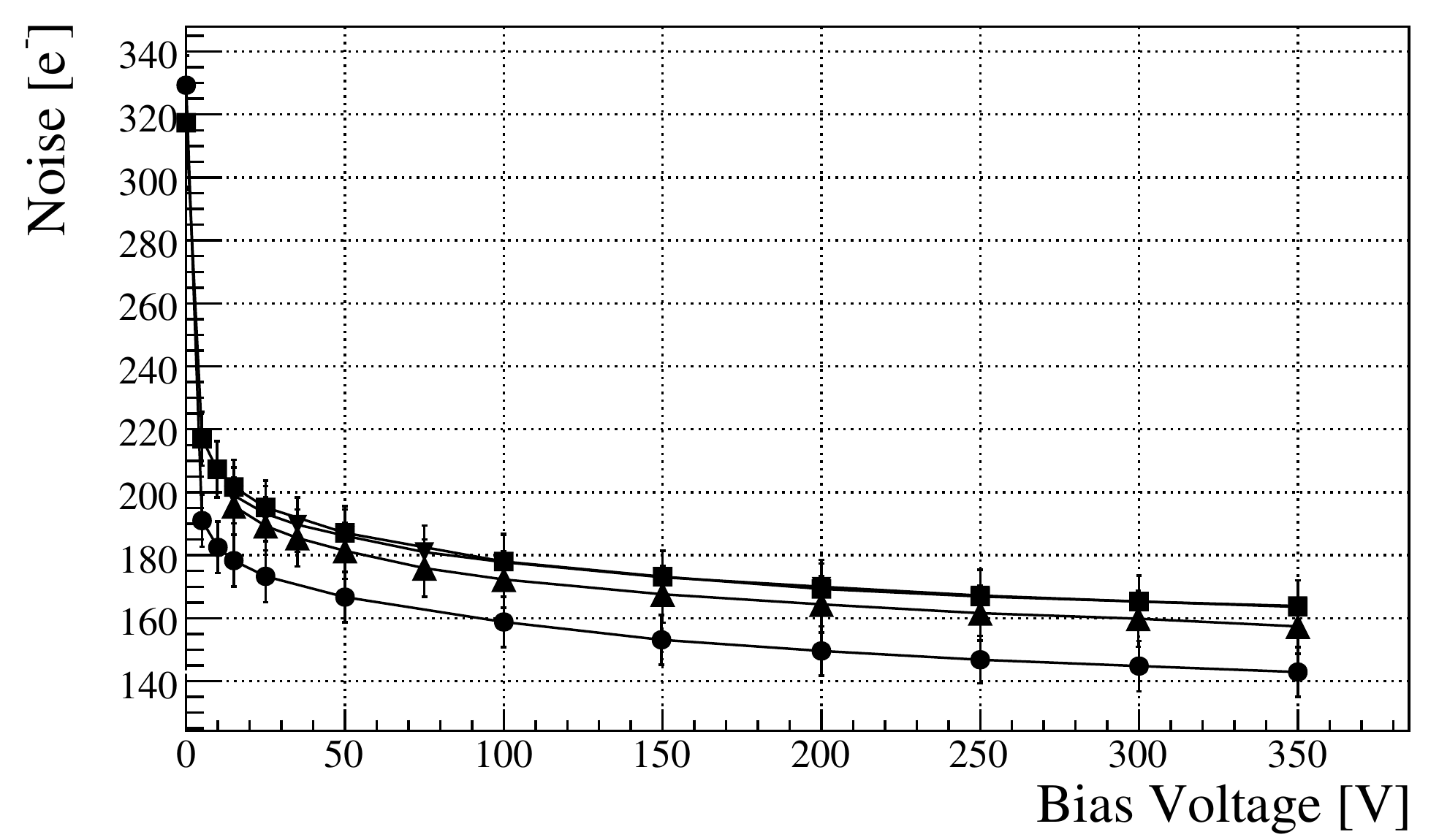}
\caption{Noise for four devices before irradiation where a bias scan was done for noise determination.}
\label{fig:Noise}
\end{figure}

The SCMs were tuned to a threshold of 3200\,\ke{} and a time over threshold (TOT) of 60 LHC-bunch-crossings for a reference signal of 20\,\ke. These tunings are comparable over all tuned devices and have a tuned threshold dispersion of (24.1$\pm$3.0)\,\e{} (USBPix) and (31.4$\pm$1.5)\,\e{} (TurboDAQ) which is comparable to the corresponding values for the current standard n-in-n technology used in the ATLAS detector \cite{pixelelectronics,Andreazza2004357}. The noise for normal pixels is (170.8$\pm$16.5)\,\e{} at 150\,V and thus comparable to the current ATLAS sensor which exhibits noise levels of 160\,$e^{-}$ \cite{pixelelectronics,Andreazza2004357}. For higher bias voltages the noise tends to decrease (Fig.\,\ref{fig:Noise}). For voltages below depletion the noise increases sharply due to the increasing capacitance of the sensor.

\subsubsection{Source Scans}
\begin{figure}[htb]
\centering
\includegraphics[angle=90,scale=0.4]{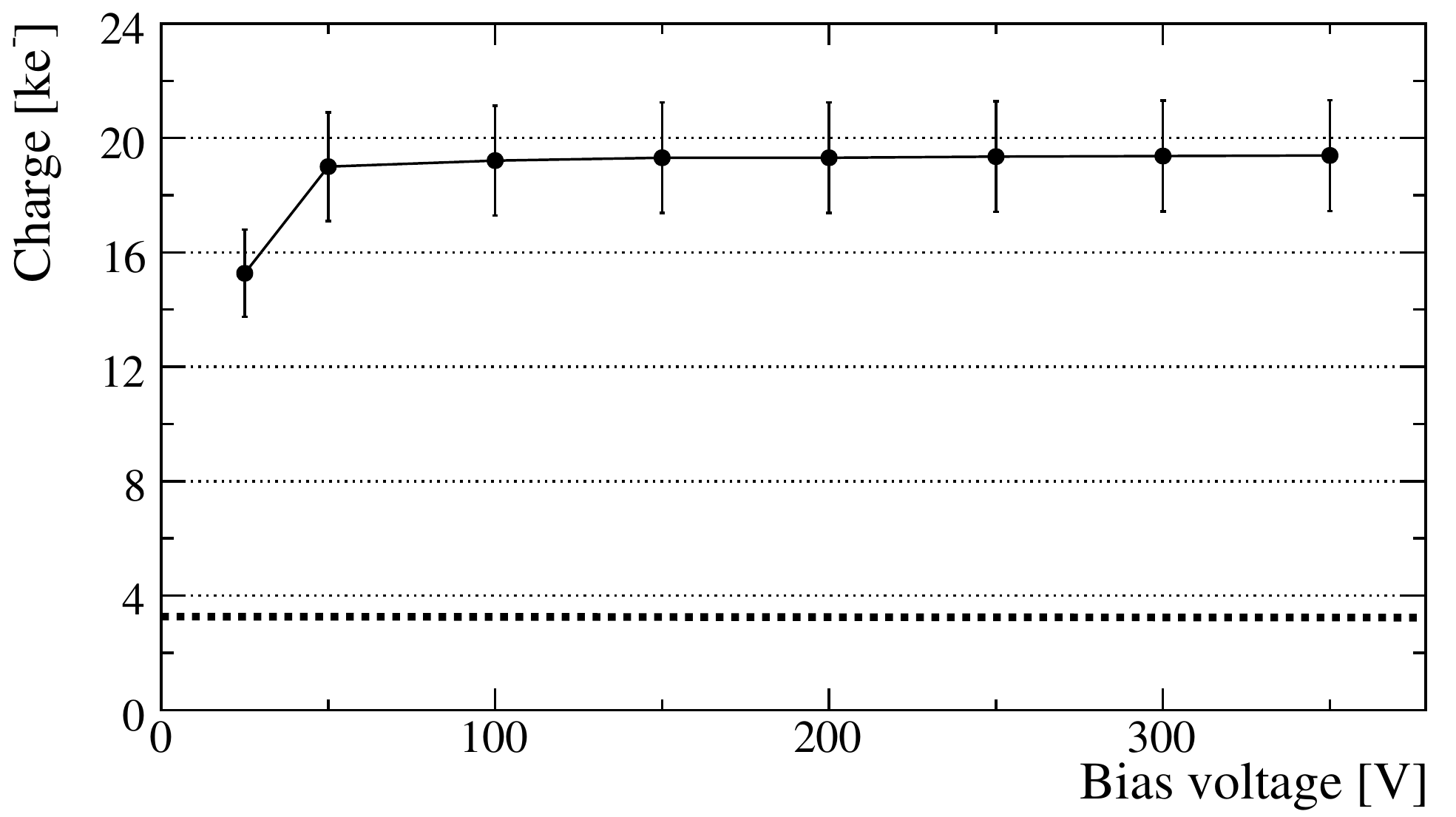}
\caption{MPV for different bias voltages between 25\,V and 350\,V for an unirradiated SCM. The dotted line indicates the threshold at 3.2\,\ke. The uncertainties account only for TOT to charge calibration. All cluster sizes are used.}
\label{fig:MPVvsbias}
\end{figure}
Source scans were conducted using a $^{90}$Sr and an $^{241}$Am source using the TurboDAQ setup of the CERN ATLAS Pixel group. 

For the $^{90}$Sr scan a collimator was used, which lead to an illumination of $10\times5$\,pixels in the central region of the SCM. At a bias voltage of 150\,V the most probable value (MPV), resulting from a fitted Landau distribution convoluted with a Gaussian, for the collected charge is 19.3\,\ke. Uncertainties arise on the one hand from the calibration of the charge to the TOT and on the other hand from the fact that charges lower than 3200\,\e{} are below threshold and thus lost. The first effect, estimated to be within 10\,\%, can modify the MPV values in both directions while the second one can only decrease the measured charge. Nonetheless the measured charge is well above threshold and thus allowing for an efficient tracking. At 150\,V ($35.9\pm1.6$)\,\% of the clusters are consisting of one and ($51.5\pm0.8$)\,\% are consisting of two pixels. For voltages above 150\,V the amount of collected charge over all cluster sizes is almost constant as shown in Fig.\,\ref{fig:MPVvsbias}. Below the depletion voltage the collected charge decreases. The uncertainties shown in Fig.\,\ref{fig:MPVvsbias} only account for the uncertainty arising from the TOT calibration.

\begin{figure}[ht]
\centering
\includegraphics[angle=90,scale=0.4]{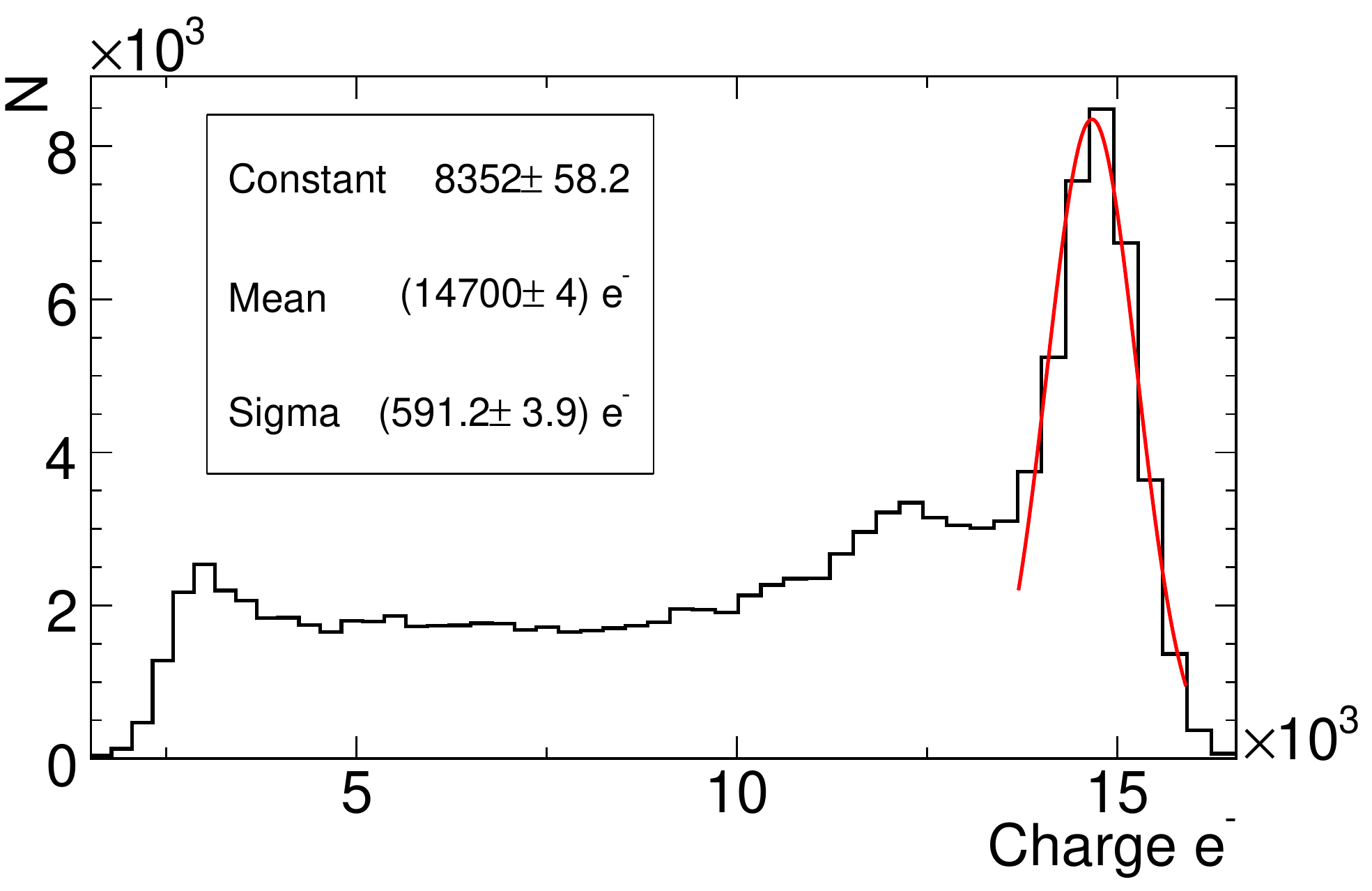}
\caption{Spectrum of an $^{241}$Am source scan at 350\,V for an unirradiated SCM. The peak region is fitted with a Gaussian.}
\label{fig:AmScan350}
\end{figure}

For a charge collection scan with an $^{241}$Am source a spectrum at 350\,V is shown in Fig.~\ref{fig:AmScan350}. A Gaussian fit to the peak resulting from the 59.5\,keV $\gamma$-line, which is expected at 16.6\,\ke, yields 14.7\,\ke. The average of the charge collection measurements over all the SCMs analysed performed with an $^{241}$Am source at bias voltages above 150\,V results in (14.2$\pm$0.7)\,\ke.

\subsection{Irradiated samples}
\label{sec:radiated}
Two SCMs were irradiated to a fluence of $10^{15}$\,\neqcm{} each. One using protons with an energy of 25\,MeV at the cyclotron of the Karlsruhe Institute of Technology (KIT) and one using reactor neutrons at the Jo\v{z}ef-Stefan-Institut in Ljubljana. Additionally three bare sensors were irradiated at CERN PS to fluences of $6.2\cdot10^{14}$\,\neqcm, $2.48\cdot10^{15}$\,\neqcm, and $4.34\cdot10^{14}$\,\neqcm.

\begin{figure}[hbt]
\centering
\includegraphics[angle=90,scale=0.4]{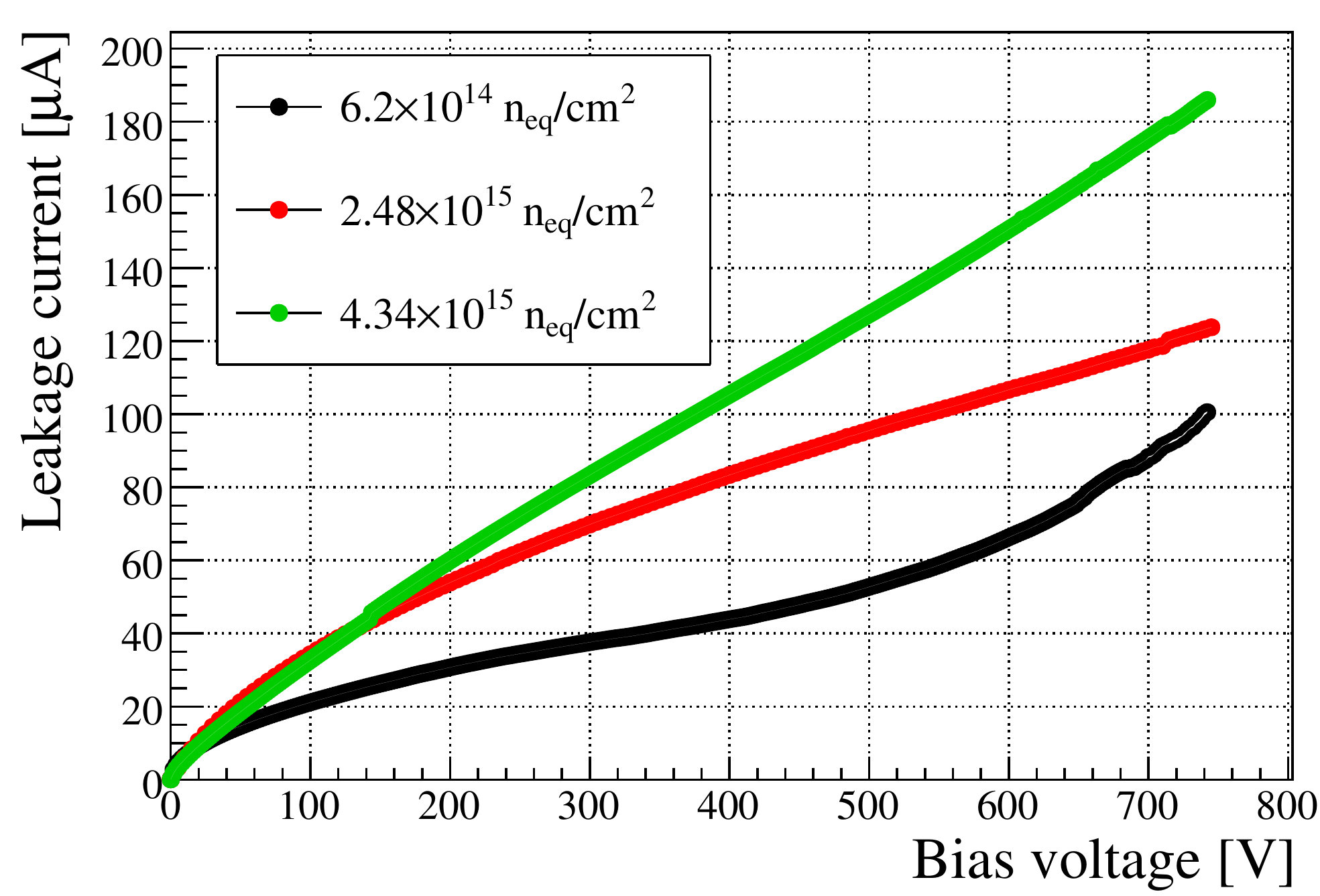}
\caption{IV characteristics for the bare sensors irradiated at CERN PS measured at -10\,$^\circ$C.}
\label{fig:BareSensors}
\end{figure}

As expected the breakdown voltage of the irradiated bare sensors shifts to higher values and are in excess of 750\,V. The leakage currents are lower than 200\,$\mu$A when measured at -10\,$^\circ$C. The IV characteristic is shown in Fig.\,\ref{fig:BareSensors}.

For the SCMs the thresholds and charges were tuned after irradiation and show a homogeneous performance, with a threshold dispersion of 25.64\,\e{} for the proton irradiated sample, which was tuned with USBPix, and of 29.64\,\e{} for the neutron irradiated sample, which was tuned with TurboDAQ.

Bias voltages up to 700\,V were applied to the devices and no sparks were seen, while stable operation for several days. The design goal for IBL is 1000\,V \cite{IBL-TDR}. To apply higher voltages the boards on which the devices are mounted need to be altered. 
\begin{figure}[h!t]
\centering
\subfigure[]{
\includegraphics[angle=90,scale=0.4]{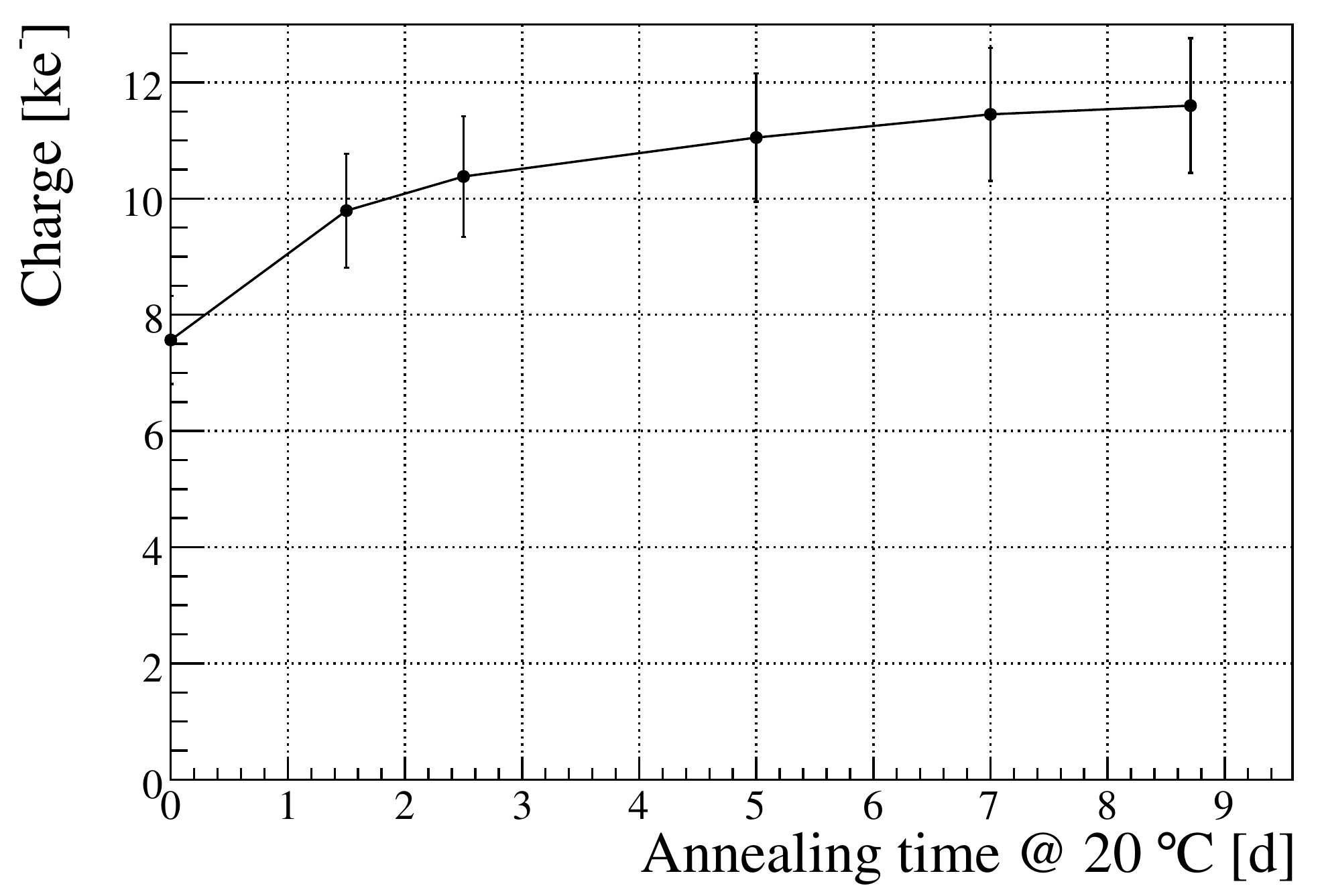}
\label{fig:Anneal}
}
\subfigure[]{
\includegraphics[angle=90,scale=0.4]{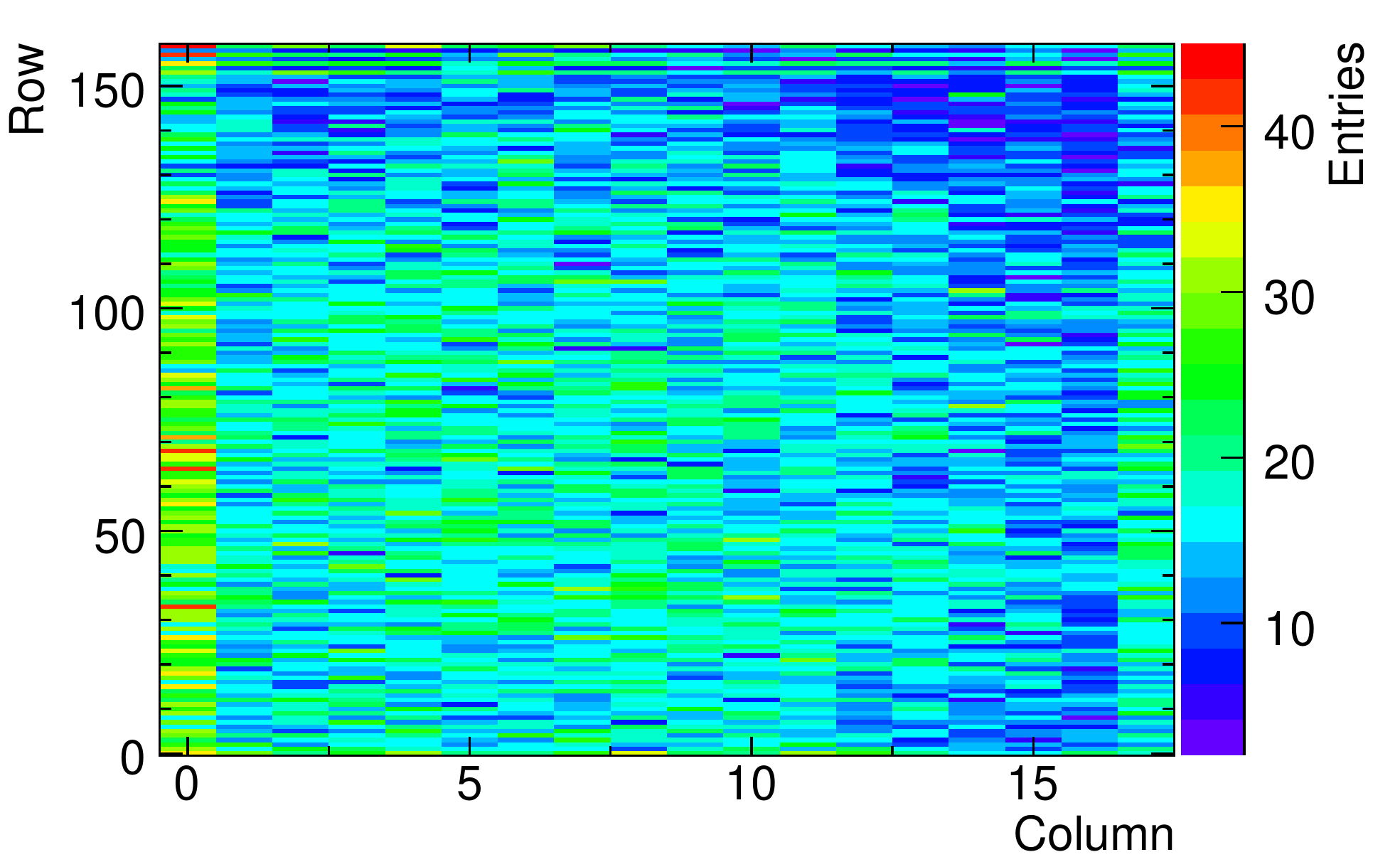}
\label{fig:anneal5hitmap}
}
\subfigure[]{
\includegraphics[angle=90,scale=0.4]{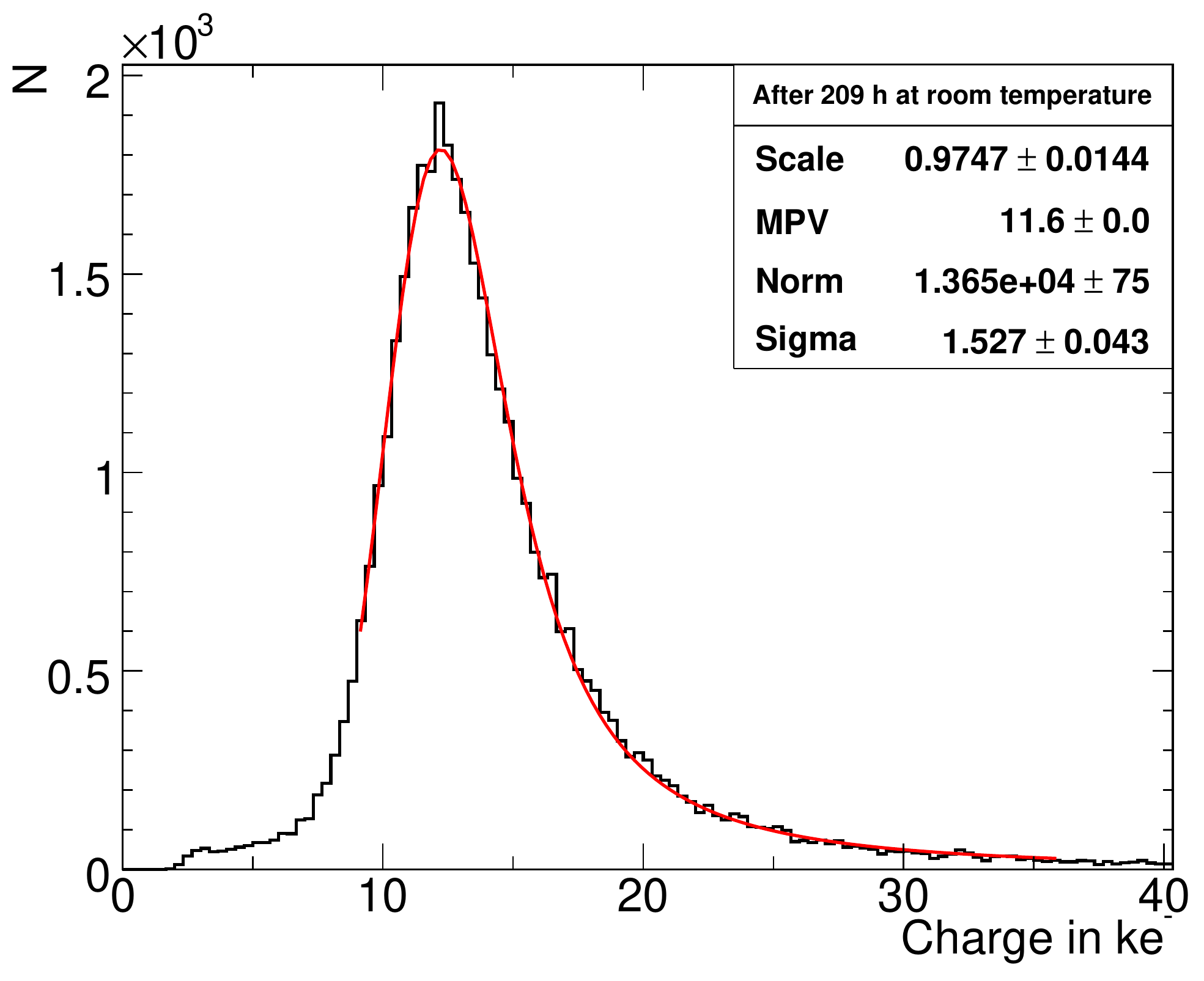}
\label{fig:anneal5landau}
}
\label{fig:subfigureExample}
\caption{\subref{fig:Anneal} Evolution of MPV over the beneficial annealing period of 209\,h at room temperature for the proton irradiated SCM. The uncertainties account only for the TOT to charge calibration. For the last annealing step a hitmap \subref{fig:anneal5hitmap} and the corresponding charge distribution \subref{fig:anneal5landau} is shown. In \subref{fig:Anneal} and  \subref{fig:anneal5landau} all cluster sizes are used.}
\end{figure}

$^{90}$Sr source scans at different bias voltages up to 550\,V for the proton irradiated sample exhibited no saturation for the collected charge, in accordance with the expected higher depletion voltage of this device. At 550\,V the most probable value of the fitted Landau function convoluted with a Gaussian is 7.6\,\ke{} directly after irradiation ($\approx$1\,h at room temperature). Over a period of almost nine days the sensor was kept at room temperature of 20\,$^\circ$C and measured after different time intervals to investigate the beneficial annealing behaviour. After this period the MPV was 11.6\,\ke at 550\,V. The full evolution is shown in Fig.\,\ref{fig:Anneal}. As an example the hit map and the charge distribution for the last annealing step are shown in Fig.\,\ref{fig:anneal5hitmap} and Fig.\,\ref{fig:anneal5landau}, respectively. In the hitmap the spot of the source can be seen in the centre shifted slightly to the left and below the centre of the sensor. The number of hits is 50\,\% higher for column 0 and 17, since these are the long pixels. The tail around 5\,\ke{} on the left of the Landau distribution (Fig.\,\ref{fig:anneal5landau}) is resulting from the not directly conected n+ implant in the bias dot region (cf. testbeam section).

\section{Testbeam Studies}
\label{sec:testbeam}
Testbeam\footnote{The PPS testbeam group has the following members: J.~Weingarten -- M.~Benoit, Ch.~Gallrapp, M.~George, S.~Grinstein, Z.~Janoska, J.~Jentsch, A.~La~Rosa, S.~Libov, D.~M\"unstermann, G.~Piacquadio, B.~Ristic, I.~Rubinsky, A.~Rummler, D.~Sutherland, G.~Troska, S.~Tsiskaridze, P.~Weigell, T.~Wittig} studies of two not irradiated SCMs were conducted at the CERN SPS/H6 beamline with 120\,GeV pions using the EUDET telescope \cite{eudet} for tracking. Due to space constraints one of the two analysed devices was placed in the middle of the six telescope planes, while the second one had to be placed behind the telescope. The resolution between the telescope planes is 3\,\mum. Behind the telescope it is expected to be better than 5\,\mum.  The device between the telescope planes is already aligned. The processing of the data for the device behind the telescope is still in progress and thus it will not be discussed in detail in these proceedings. Preliminary analyses suggest that the results for this device are comparable to the ones presented here.

Further information on basic testbeam analyses can be found in \cite{WeingartenTB}.

\subsection{Testbeam Analysis}
For the analysis of the testbeam data a fiducial region spanning 14 columns and 118 rows in the centre of the sensor was used. Finally, all analysed pixels were overlayed. To match the tracks to reconstructed clusters a search window centred at each pixel and extending $\pm$400\,\mum{} along the long pixel side and $\pm$150\,\mum{} along the short pixel side is used. 
\begin{figure*}[htb]
\centering
\centering
\subfigure[]{
\includegraphics[angle=90,scale=0.93]{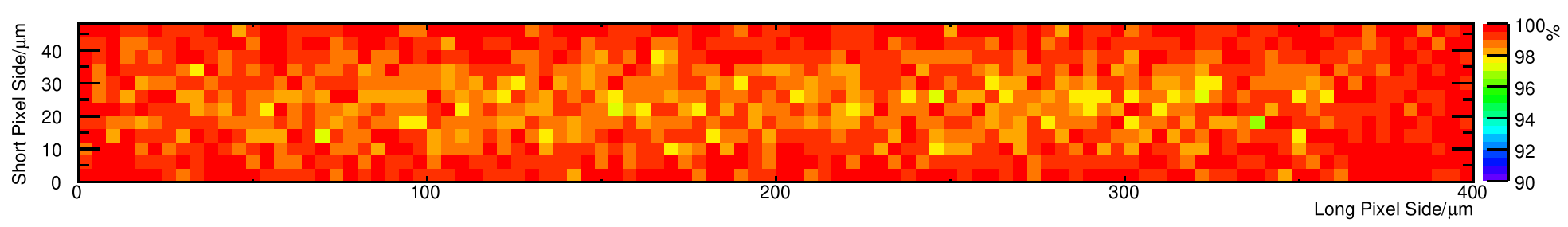}
\label{fig:tackeff}
}
\subfigure[]{
\includegraphics[angle=90,scale=0.93]{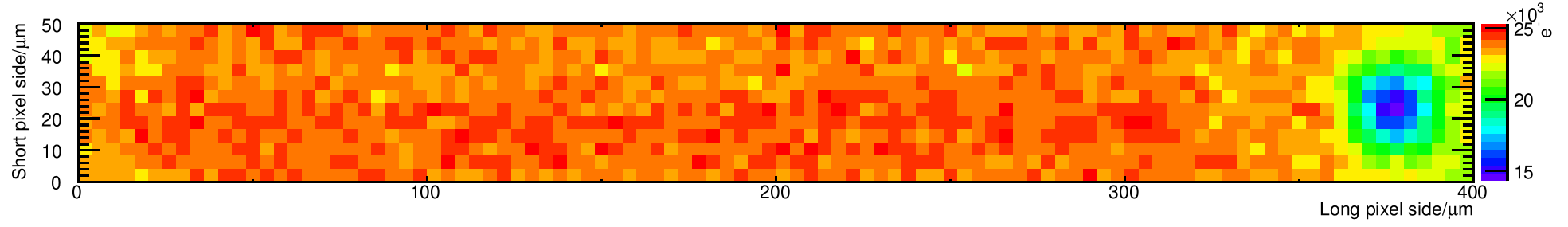}
\label{fig:ChargeMap}
}
\caption{Map of mean tracking efficiency \subref{fig:tackeff} and of mean collected charge \subref{fig:ChargeMap} by track position within one pixel. In Fig.\,\ref{fig:ChargeMap} the bias dot is visible clearly on the right hand side.}
\end{figure*}

\subsubsection{Tracking Efficiency}
The tracking efficiency for the device, defined by the ratio of the number of matched clusters to the number of all tracks reconstructed with the telescope is found to be 99.3\,\%. In Fig.\,\ref{fig:tackeff} the mean tracking efficiency by track position within a pixel is shown.

\subsubsection{Charge Collection}
The mean charge---differently from the MPV used in the previous figures---for all cluster sizes is shown in Fig.\,\ref{fig:ChargeMap}. For the main region of the chip the collected charge is constant at a level of 24\,\ke{}. In the region of the punch through biasing, where the central dot of the n+ implantation is not directly connected to the pixel structure, the mean is about 10\,\ke{} lower than in the central region of the pixel, but still well above threshold. In a not spacial resolved charge distribution like in Fig.\,\ref{fig:anneal5landau} this region results in a plateau at lower charge values. A similar effect is seen in the left corners of Fig.\,\ref{fig:ChargeMap}. Here the charge collection is slightly decreased, due to the higher charge sharing probability (three neighbouring pixels) possibly causing one pixel to be below threshold.

\section{Conclusion}
\label{sec:conclusion}
For the first time working n-in-p based pixel detector SCMs were characterised in the lab and in a testbeam. Also for the first time, irradiated samples of such devices were studied. 

As expected the flip-chipping did not deteriorate the sensor performance. The stable operation of 700\,V for several days indicates that the risk of sparks is low if BCB is applied. The tuned  threshold, its dispersion as well as the noise is comparable with the standard n-in-n technology.

An excellent signal to threshold ratio above $19.3/3.2$  was determined. After irradiation up to $10^{15}$\,\neqcm{} the ratio is still sufficient even for not fully depleted sensors. The beneficial annealing behaviour over nine days at room temperature was as expected.

The SCMs performed well in source scans with $^{90}$Sr and $^{241}$Am as well and showed excellent tracking and charge collections efficiencies in a pion testbeam. 

\section{Acknowledgements}
\label{sec:acknowledgment}
This work has been partially performed in the framework of the CERN RD50 Collaboration. The authors thank A.~Dierlamm (KIT), V.~Cindro (Jo\v{z}ef-Stefan-Institut) and M.~Glaser (CERN PS) for the sensor irradiation. Part of the irradiation were performed within the framework of the Helmholtz alliance. The testbeam was supported by EUDET.


\begin{thebibliography}{99}
\bibitem{SLHC} M. Lamont, ``LHC near and medium term prospects'', Proceeding of Physics at the LHC 2010, Hamburg, 7-12 June 2010.
\bibitem{Dawson}I. Dawson and L.Nicolas, these proceedings
\bibitem{Wedenig1999497}R. Wedenig et al., ``CVD diamond pixel detectors for LHC experiments'', Nucl. Phys. B - Proceedings Supplements, Vol. 78(1-3) (1999) 497--504
\bibitem{Parker3D}S. I. Parker et al., ``3D - A proposed new architecture for solid-state radiation detectors'', NIM A, Vol. 395 (1997) 328.
\bibitem{GarciaSciveres2010} M. Garcia-Sciveres et al., ``The FE-I4 pixel readout integrated circuit'', NIM A (2010) (In press)
\bibitem{Peric2006178}I. Peric et al. ``The FEI3 readout chip for the ATLAS pixel detector'', NIM A, Vol. 565, No. 1 (2010), 178--187
\bibitem{IBL-TDR} M. Capeans et al., ``ATLAS Insertable B-Layer Technical Design Report'', CERN-LHCC-2010-013, Geneva Sep. 2010
\bibitem{thinning} L. Andricek et al., IEEE Trans. Nucl. Sci., Vol. 51, No. 3. (2004), 1117.
\bibitem{eudet} A. Bulgheroni, "Results from the EUDET telescope with high resolution planes'', NIM A, Vol. 623, No. 1 (2010), 399--401
\bibitem{WeingartenTB} J. Weingarten et al., ``The PPS testbeam: performance and first analysis results'', (in preparation)
\bibitem{pixelelectronics} G.~Aad, et al. ``ATLAS pixel detector electronics and sensors'', JINST, Vol. 3(7) (2008), P07007
\bibitem{Andreazza2004357} A.~Andreazza, ``Developments of the ATLAS pixel detector'', NIM A, Vol. 535, No. 1--2 (2010), 357--361
\end{thebibliography}
\end{document}